\documentclass[twocolumn,showpacs,amsmath,amssymb,pra,superscriptaddress,floatfi]{revtex4}

\usepackage{graphicx}
\usepackage{epsfig}
\usepackage{dcolumn}
\usepackage{bm}
\newcommand{\bg}[1]{\mbox{\boldmath$#1$}} 
\usepackage{xcolor}
\definecolor{red}{rgb}{0.7,0,0}
\definecolor{green}{rgb}{0.,0.35,0.}
\definecolor{blue}{rgb}{0.2,0.2,0.7} 
\definecolor{black}{rgb}{0.15,0.15,.15}

\begin{document}

\title{Mesoscopic dipolar quantum crystals}
\pacs{67.85.-d, 03.75.-b, 32.80.Ee, 34.20.Cf}
\author{Massimo Boninsegni}
\affiliation{Department of Physics, University of Alberta, Edmonton, Alberta, Canada T6G 2J1}
\email{m.boninsegni@ualberta.ca}
\begin{abstract}
The ground state of a two-dimensional, harmonically confined mesoscopic assembly of up to thirty polar molecules is 
studied by computer simulations. As the strength of the confining trap is increased, clusters evolve from superfluid, to supersolid, to insulating crystals.
For strong confinement, the  crystalline structure can be predicted based on classical energetics. However, clusters of specific numbers of particles (i.e., $N$=12 and $N$=19) display a {\it non-classical crystalline structure}, stabilized by  quantum effects, in an intermediate range of confinement strength. In these cases, coexistence of quantum and classical crystalline configurations is observed at finite temperature.
\end{abstract} \maketitle
\section{Introduction}
One of the most remarkable recent experimental achievements is 
the preparation of cold ensembles of dipolar particles. Gases of ultracold
ground state magnetic atoms have been shown to display dipole-dominated dynamics in the quantum regime~\cite{Lahaye2007,Billy2012,Pasquiou2012,Lu2012,Aikawa2012}. Huge electric dipole moments, of up to thousands of Debye,
are present in  highly excited Rydberg states of Alkali atoms~\cite{GallagherBook,Saffman,Comparat}, and experiments are under way to demonstrate and control interactions in these systems~\cite{RydbergExp}. Strong correlations with ground state atoms can be obtained 
by weakly admixing with laser light these Rydberg states \cite{Santos,guido,Henkel2010,Cinti2010}, at the price of finite heating and losses due to spontaneous emission~\cite{Glaetzle2012}.
Polar molecules prepared in the electronic and rovibrational ground state~\cite{PolMolBook,Doyle2004,Carr2009,Jin2012,PolMolExp} combine the stability of ground state particles with dipole moments of up to a few Debye, leading to large dipolar interactions. Experiments are well under way to exploit these interactions in combination with reduced trapping geometries, to add collisional stability~\cite{deMiranda2011,Chotia2012}. This
opens the door to the study of strongly correlated quantum phases with
designed long range interactions~\cite{Lahaye2007,Baranov2012}, e.g., for dipolar Bose systems the density-driven superfluid-crystal quantum phase transition in 2D ~\cite{Buechler07,Astrakharchik2007,Mora2007}.
\\ \indent
Finite, confined dipolar assemblies are  worthy of investigation for a number of reasons, the most obvious being the study of the evolution of the physical properties of the system as its size is increased, approaching the bulk limit \cite{castr,cinti}. For example, recent theoretical work with Fermionic molecules~\cite{Bruun2008,Baranov2011,Parish2012,Matveeva2012} has analyzed the effects of quantum statistics on transitions between Wigner-type states in traps~\cite{Cremon2010}. In this work, we focus our attention on quasi-two-dimensional, harmonically confined dipolar systems of mesoscopic size, comprising a relatively small number of particles. One of the fundamental issues that one  may address, is the occurrence of superfluidity in a  cluster, until now mostly explored in the context of helium \cite{sindzingre} and hydrogen \cite{sindzingre2,noi,noi2} droplets.\\ \indent
Based on results of first principle quantum simulations, an intriguing deviation from bulk behaviour that occurs in a dipolar system of small size is predicted here, namely the occurrence of {\it non-classical} ground state crystalline arrangements, in an intermediate range of confinement strength. These configurations differ from the classical arrangement by the number of particles in the inner shell of the crystal, and are stabilized by the energetic contribution of  zero-point motion.  To our knowledge, this is the first demonstration of such a physical effect, which can be observed experimentally in mesoscopic dipolar clusters of polar molecules. It is observed for clusters of specific sizes, namely $N$=12 and $N=19$, for $N \le 30$. This effect takes place in the mesoscopic crystalline phase, in which quantum-mechanical exchanges of indistinguishable particles are stroingly suppressed; it is therefore expected to take place irrespective of the quantum statistics of the particles.
\\ \indent
We describe the mathematical model of the system utilized here in the next section; we then briefly review the computational methodology utilized here, and illustrate the results of the calculations and main physical conclusions in the following sections.
\section{Model}
We consider a setup where $N$ dipolar particles of mass $m$ are confined to a 2D plane by applying a strong transverse trapping field~\cite{Buechler07}, e.g a 1D optical lattice. Dipole moments are aligned perpendicular to the plane, with a DC induced dipole moment $d\equiv \sqrt{D} $. We assume an additional in-plane parabolic trap with frequency $\omega$, as realized by a magnetic dipole trap, or a single site of a large spacing optical lattice. For pure dipolar interactions, the many-body Hamiltonian in dimensionless form is given by
\begin{eqnarray}\label{eq:eqHamRes}
{\hat H}=\frac{1}{2}\sum_{i=1}^{N} \left[-\nabla_i^2+{\Gamma\bg r}_i^2\right]
    + \sum_{i< j}\frac{1}{|{\bg r}_i-{\bg r}_j|^3},
\end{eqnarray}
where the unit of length is $a\equiv (mD/\hbar^2)$, that of energy is $\epsilon_\circ \equiv (D/a^3)\equiv(\hbar^2/ma^2)$, and where $\Gamma\equiv (1/\xi^4)$, $\xi\equiv\sqrt{\hbar/(m\omega a^2)}$ being the characteristic (dimensionless) confining length of the parabolic trap. $\Gamma$ plays the role of control parameter here, as a greater value of $\Gamma$ means that the trap is compressed, and the particle density correspondingly increased.  We assume for definiteness that particles obey Bose statistics, and all of the numerical results presented here 
are obtained with this assumption; however, the conclusions of this study pertaining to the crystalline phases
are {independent} of quantum statistics. \\ \indent
The basic ground state physics of the system described by (\ref {eq:eqHamRes}) has been characterized in previous works \cite{castr,cinti}; generally speaking, it mimics qualitatively that of the bulk system \cite{Buechler07}. For sufficiently small $\Gamma$, the cluster is in a low-density, weakly interacting superfluid phase \cite{qualifier}. In the limit of strong confinement, the potential energy dominates, and the  ground state takes on the classical lowest-energy crystalline configuration,  predictable by straightforward potential energy minimization (last two terms of Eq.~\eqref{eq:eqHamRes}). \\
For a cluster comprising a sufficiently small number of particles (i.e., $N\lesssim 30$), the finite size of the system allows for the existence of several intermediate, ``supersolid"  phases \cite{guido,castr} displaying simultaneously a finite superfluid response and crystalline order \cite{h2}. These phases are not expected  to survive in the thermodynamic limit; indeed, the nature of a supersolid phase of a system of dipolar bosons in two dimensions is predicted \cite{Kivelson} to deviate significantly from that of  mesoscopic supersolids.
\\ \indent
In this manuscript we mostly  focus our attention on mesoscopic crystalline phases, in which particles are localized and consequently quantum-mechanical exchanges are suppressed. 
\section{Methodology}
We investigated the low temperature ($T\to 0$) properties of the system described by (\ref{eq:eqHamRes}) by means of computer simulations, based on the continuous-space Worm Algorithm \cite{worm,worm2,roy}. Since this technique is by now fairly well-established, and extensively described in the literature, we shall not review it here. Details of the simulation are standard. \\ \indent 
 Because we are interested in the physics of the system in the $T\to 0$ limit, we report here results corresponding to a temperature $T$ sufficiently low to regard them as essentially ground state estimates. 
A quantitative criterion to assess whether the temperature $T$ of the simulation is sufficienty low, consists of comparing it to the computed  particle mean kinetic energy $\langle K\rangle$.
For all the simulations for which we present results in Figures \ref{f2} and \ref{f4}, it is $T \lesssim 3\times 10^{-2}\ \langle K\rangle $. \\ \indent
The use of a finite temperature technique to investigate ground state physics might appear counterintuitive, considering that methods exist purposefully designed to study the ground state of a many-body system (e.g.,  Diffusion Monte Carlo). In practice, however,  finite-temperature techniques often prove superior, even to determine ground state properties (naive statements by DMC practitioners notwithstanding). This is mainly owing to the unbiasedness of finite temperature methods, which, unlike their $T$=0 counterparts, require no {\it a priori} physical input (e.g., a trial wave function). Moreover, finite temperature methods allow one to assess more easily and reliably quantities other than the energy, including off-diangonal correlations.
We compute both global and local superfluid responses, using standard methodology \cite{noi4}. All of the results presented here are extrapolated to the limit of zero imaginary time step \cite{roy}. 
\section{Results}
\begin{figure}
\includegraphics[width=0.9\columnwidth]{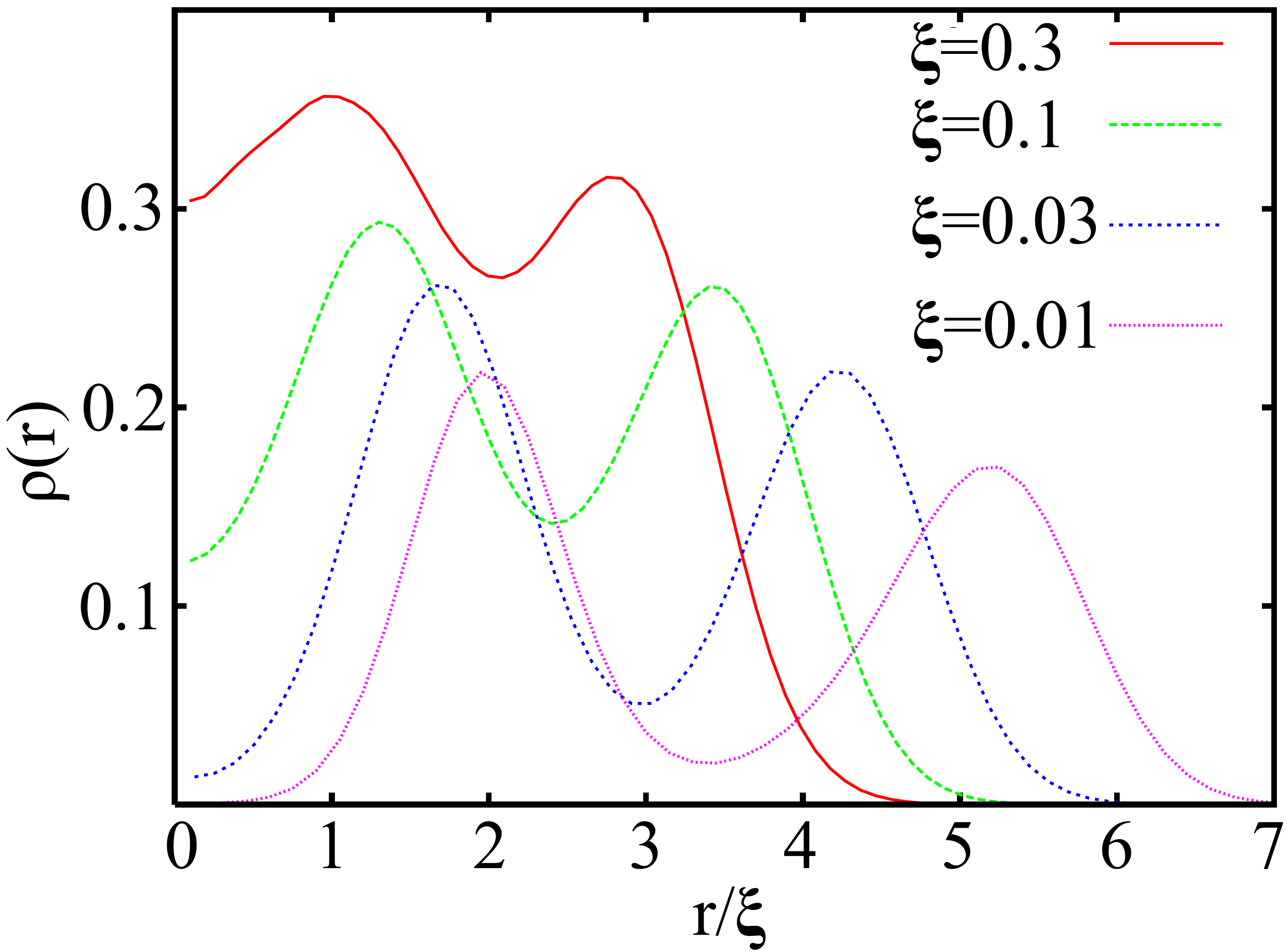}
\caption{{\it Color online}. Radial density profiles $\rho(r)$ for a two-dimensional system of $N$=12 dipolar bosons confined in a planar harmonic trap of different characteristic length $\xi$.}\label{fig:fig2}
\end{figure}
\begin{figure}
\centering
\begin{tabular}{cc}
\includegraphics[width=0.46\columnwidth]{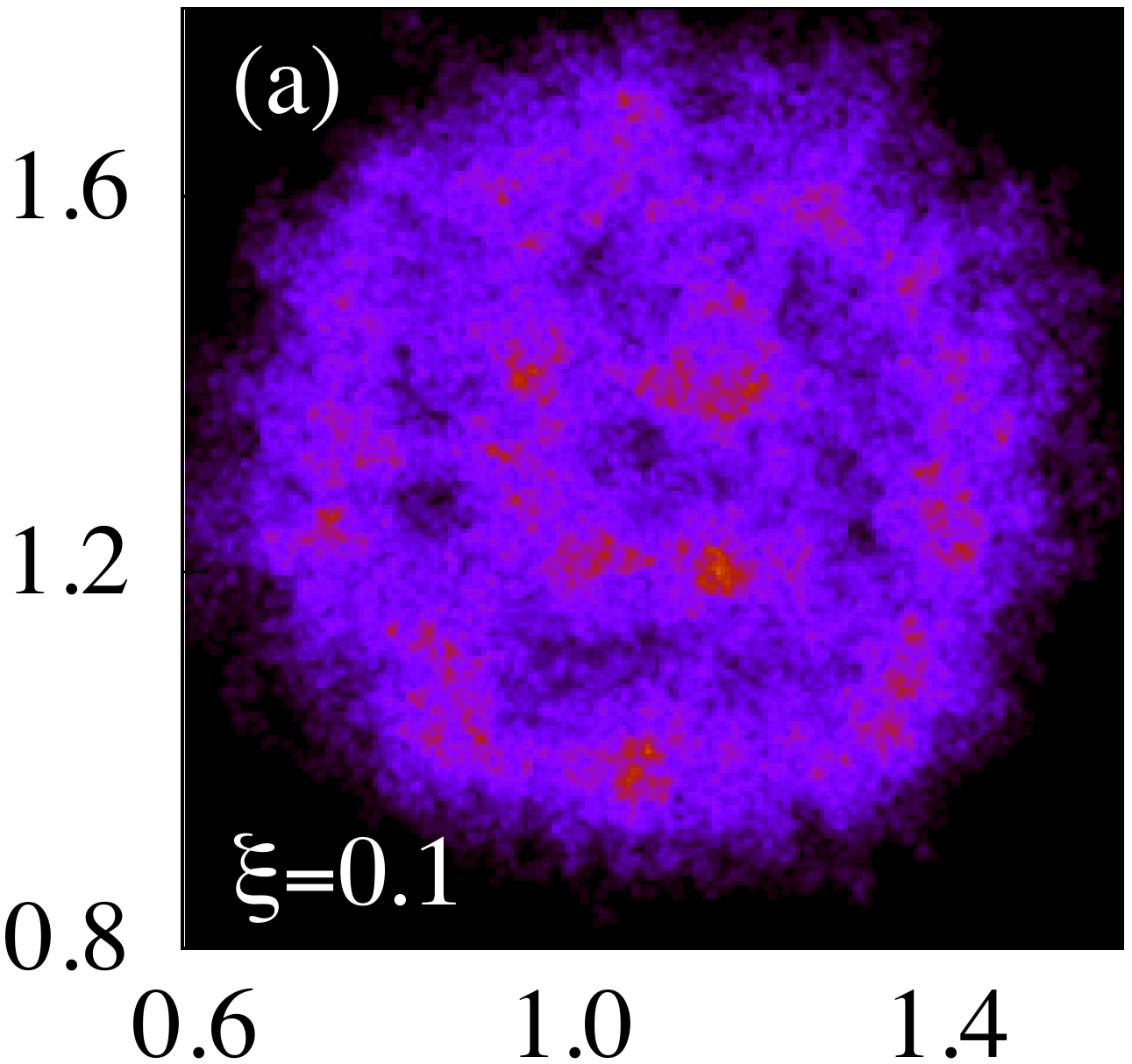}&
\includegraphics[width=0.46\columnwidth]{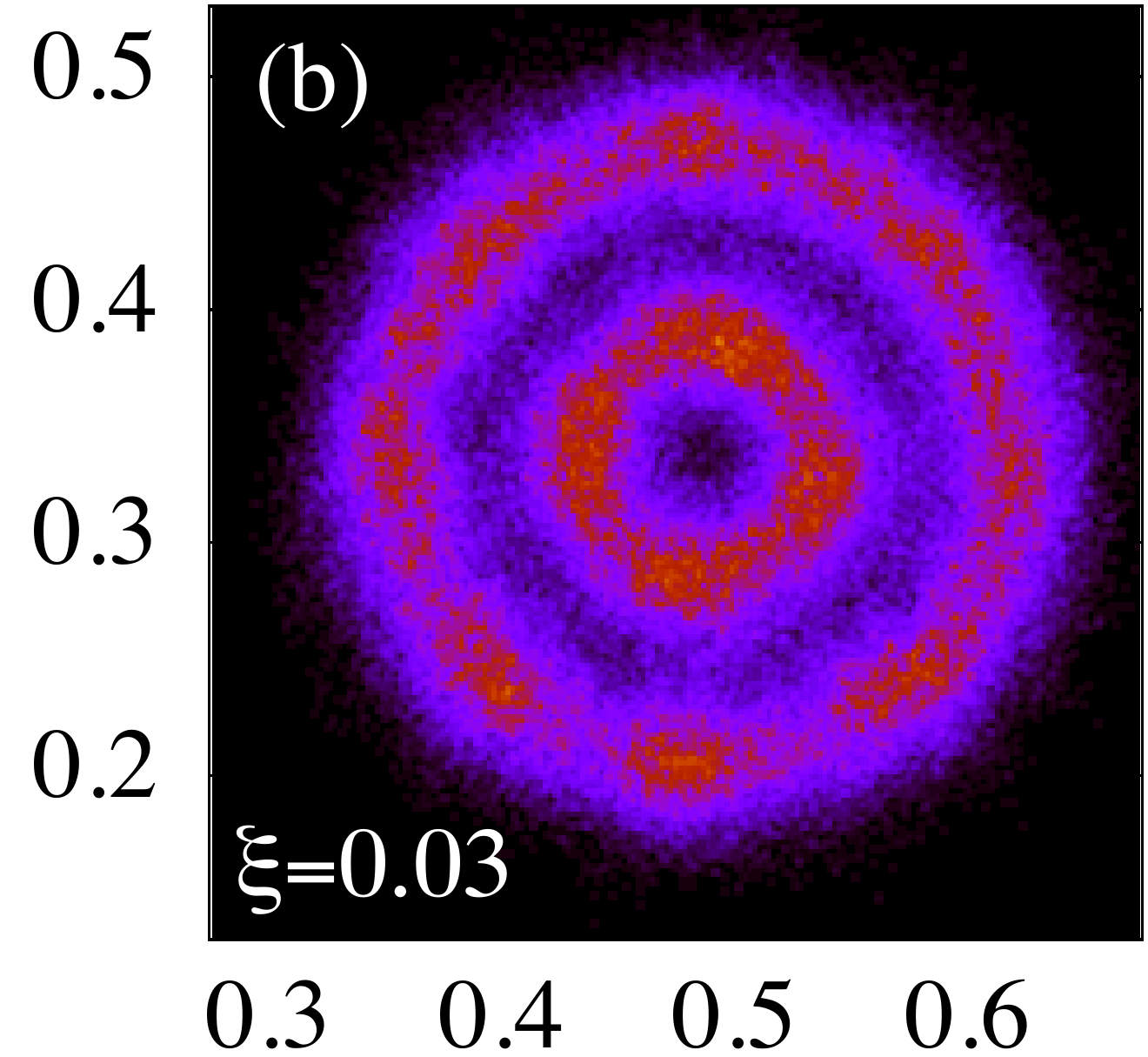}\\
\includegraphics[width=0.46\columnwidth]{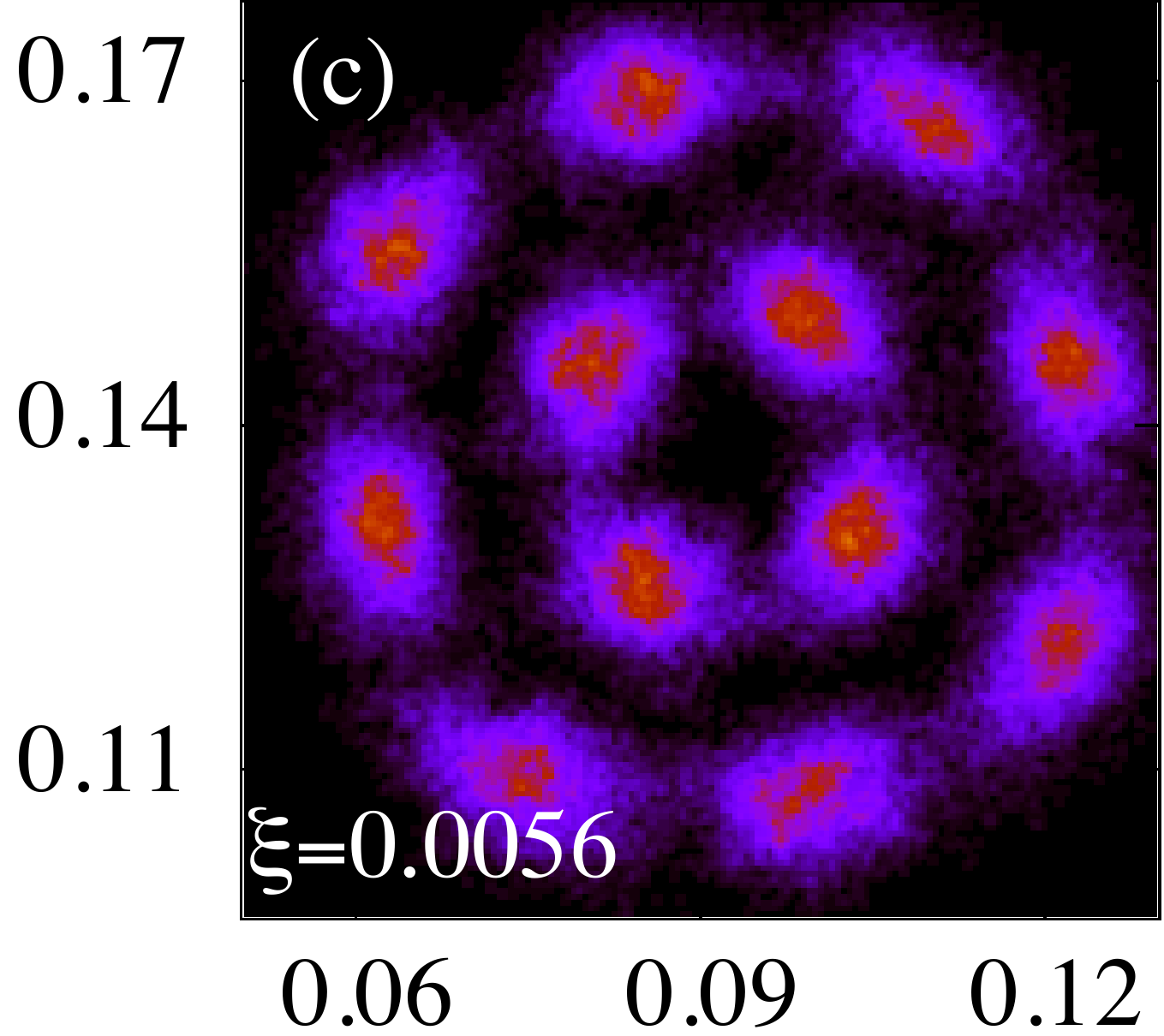}&
\includegraphics[width=0.46\columnwidth]{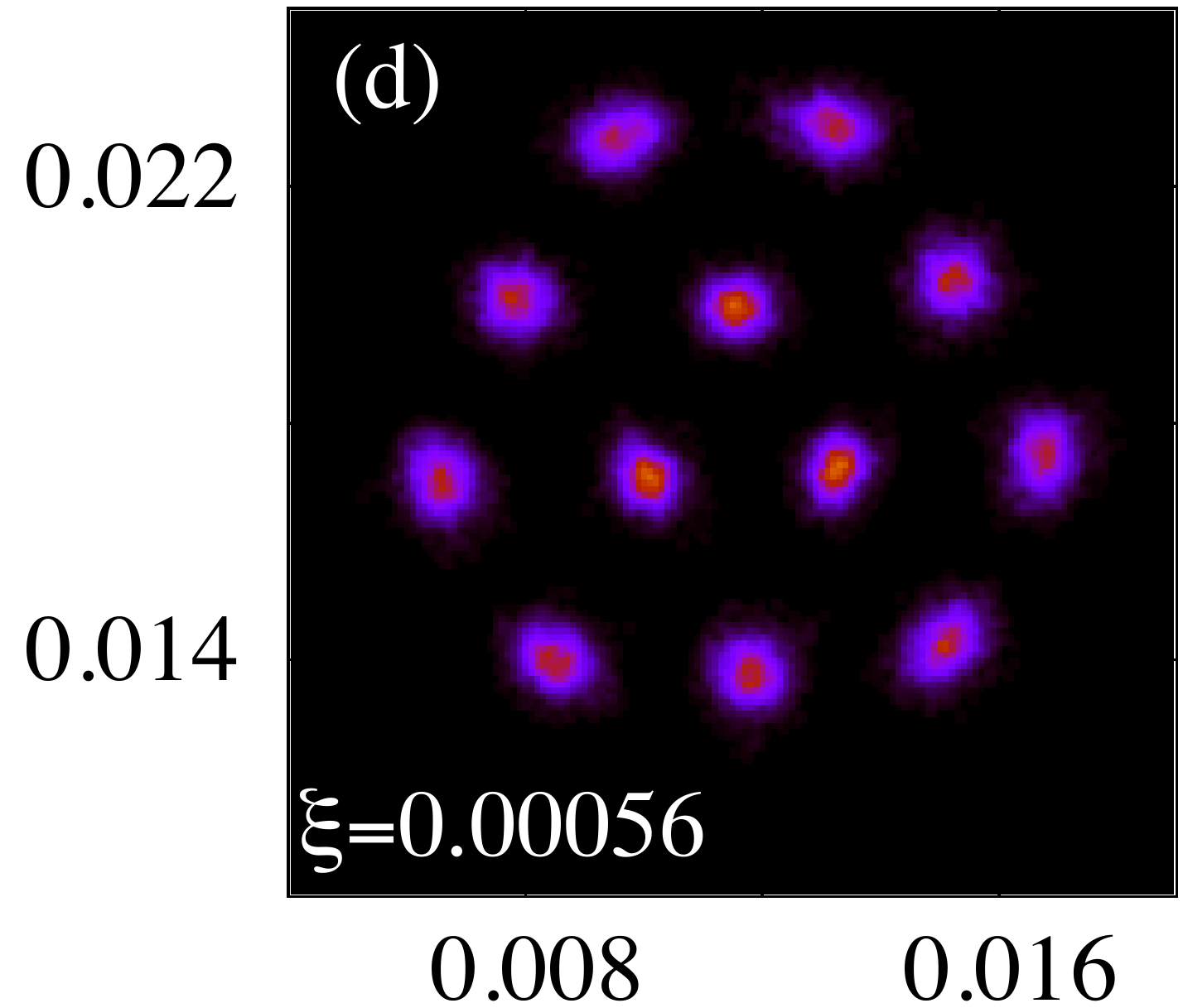}\\
\end{tabular}
\caption{{\it Color online}. Ground state density maps for a two-dimensional mesoscopic system of $N$=12 dipolar bosons confined in a harmonic trap of  varying characteristic length $\xi$. All lengths are all in units of $a$ (see text). 
Panels (a) shows a superfluid, (b) a supersolid, (c) a non-classical crystal and (d) a classical crystal. }\label{f2}
\end{figure}

Figure \ref{fig:fig2} displays radial density profiles (computed with respect to the center of the trap) for the ground state of a mesoscopic assembly of $N$=12 particles, confined in two-dimensional harmonic traps of varying strength (i.e., characteristic length). In order to facilitate the comparison, we plot the radial density $\rho(r)$ as a function of $r/\xi$. 
Figure \ref{f2} shows particle density maps obtained from statistically representative  configuration snapshots (i.e., particle world lines) for four distinct cases, corresponding to different values of the harmonic confining length $\xi$, namely $\xi=0.1, 0.03, 0.0056$ and 0.00056. By ``statistically" representative, we mean that every configuration generated in the simulation is physically equivalent to that shown in the figure; in particular, for the case in which the system takes on the crystalline arrangements shown in panels (c) and (d), every configuration in the Monte Carlo random walk only differs from that shown by a mere rotation.
\\ \indent
For a relatively weak trap (e.g., with $\xi \gtrsim 0.1$) the radial density profile is largely featureless, only displaying two broad shoulders; correspondingly, configuration snapshots (panel (a) in Fig. \ref{f2}) display a fairly uniform particle density, aside from local random fluctuations. This system enjoys fluid-like behavior, with frequent exchanges of indistinguishable (Bose) particles, leading to a robust ground state superfluid response, uniformly distrbuted throughout the cluster and approaching 100\% in the $T$=0 limit.  As confinement is rendered tighter,  two increasingly well-defined shells appear, and the system takes on solid-like configurations, wherein particles form two concentric rings, as shown in panels (b), (c) and (d) of Figure \ref{f2}. As shown in Fig.  \ref{fig:fig2}, for a trap size $\xi \lesssim 0.01$ the two shells are essentially non-overlapping. Also worth noticing is the fact that, while for weak confinement the size of the cluster is essentially related to the value of $\xi$, for tight confinement is largely determined by the repulsive interaction among dipoles.
\\ \indent
In the limit of very tight confinement ($\xi \lesssim 0.001$) the system takes on the configuration that minimizes the classical potential energy (panel (d) in Figure \ref{f2}). Here, particle localization suppresses exchanges and the superfluid density drops to zero \cite{note}. For intermediate values of $\xi$ (roughly 0.001$\lesssim\xi\lesssim$ 0.05), the competition between dipolar interactions, harmonic confinement and quantum delocalization has the effect of stabilizing mesoscopic phases with no classical counterpart, neither observed in the quantum-mechanical bulk system. For $\xi\gtrsim 0.02$, the cluster forms a ring-shaped ``supersolid" (panel (b) of Figure \ref{f2}),  whereas a non-superfluid non-classical crystal (panel (c) of Figure \ref {f2}) is stable close to $\xi\lesssim 0.001$.
\\ \indent
In the mesoscopic supersolid phase, the system forms two concentric rings, the outer (inner) ring comprising eight (four) particles. Remarkably, the superfluid response is quantitatively unaltered compared to the unmodulated superfluid phase depicted in panel (a). Specifically, the superfluid fraction is observed to approach unity in the $T\to 0$ limit. Similarly to what observed in hydrogen clusters \cite{noi4}, superfluidity in this cluster is underlain by cycles of exchanges involving particles in different rings, and the local superfluid fraction is essentially homogeneous throughout the cluster (equivalently, the behavior of the local superfluid density mimics that of the local density).
\\ \indent
Superfluidity is dramatically suppressed around $\xi\sim 0.01$. As shown in panel (c) of Figure \ref{f2}, overlap between the quantum delocalization ``clouds" associated to different particles is minimal, rare exchanges occurring only among particles in the inner ring. From the density profile shown in Figure  \ref{fig:fig2} we infer that the radius of the cluster for $\xi=0.01$ is approximately given by $7\xi$, corresponding to a value of the mean inter-particle distance $r_s$ around 0.035 (in units of $a$). In the bulk system, the superfluid-to-insulator transition is predicted \cite{Buechler07} to occur for $r_s\approx 0.056$, a value for which the mesoscopic system of interest here is in the ``supersolid" phase described above.
\begin{figure}
\centering
\begin{tabular}{cc}
\includegraphics[width=0.46\columnwidth]{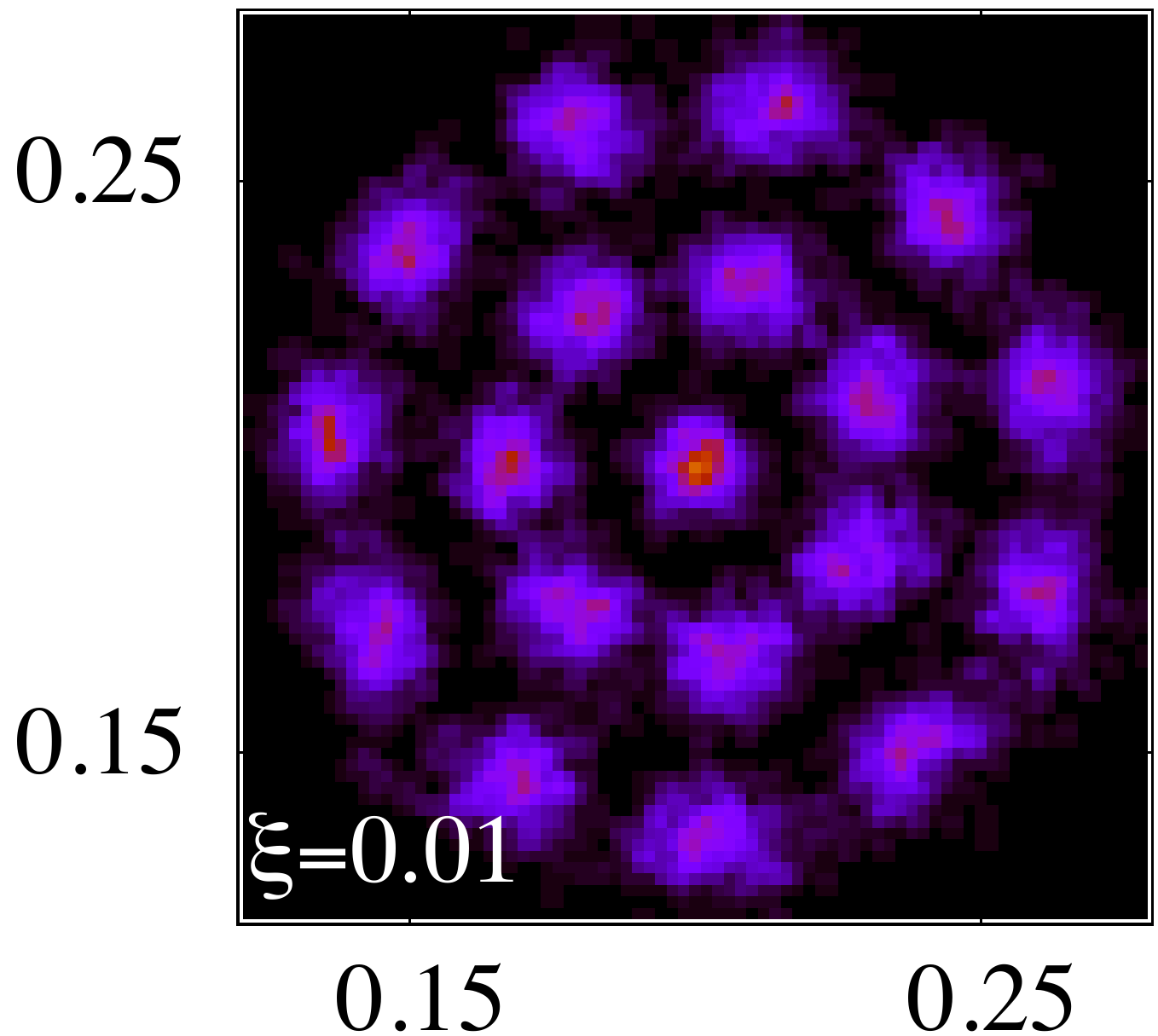}&
\includegraphics[width=0.46\columnwidth]{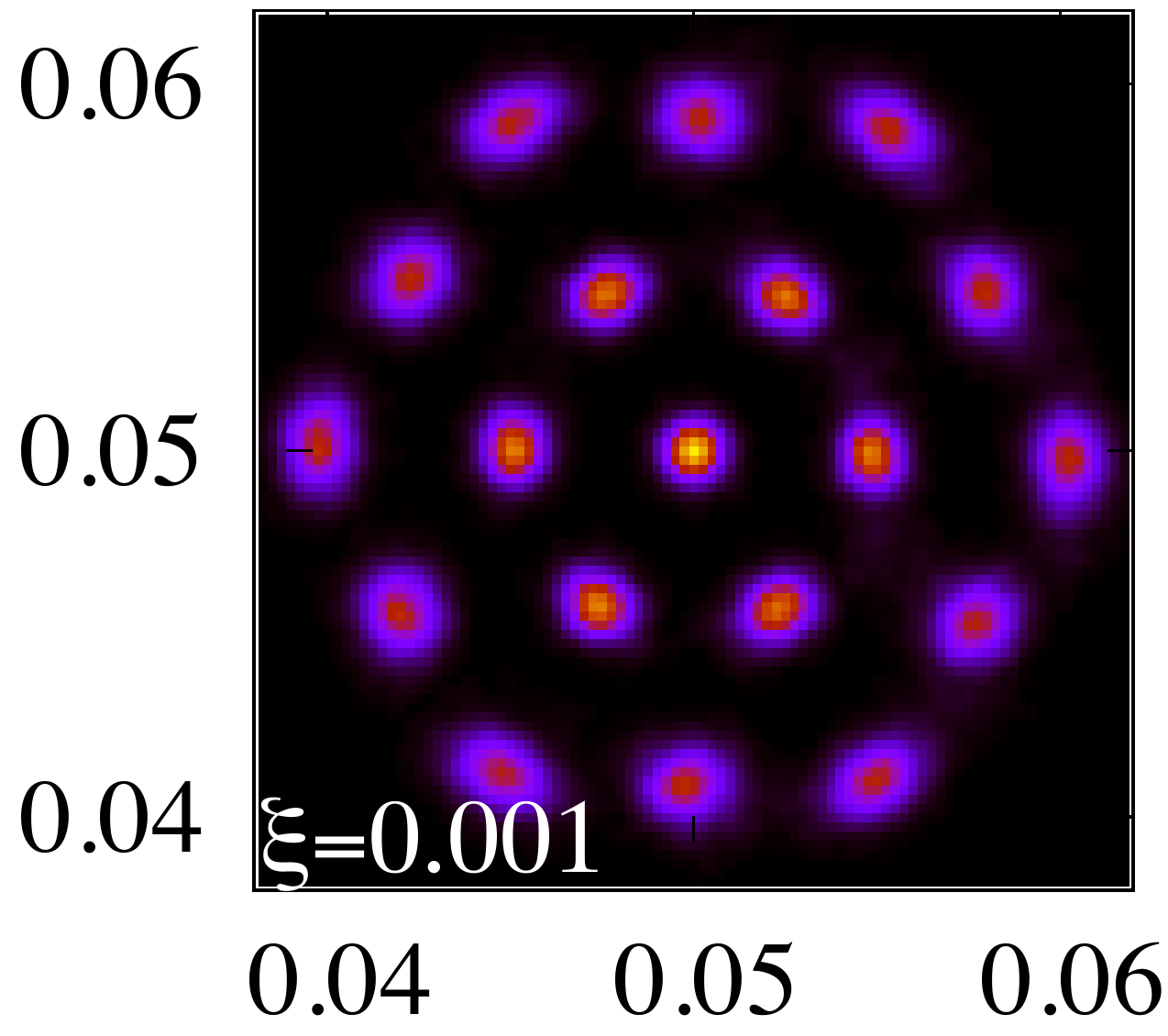}
\end{tabular}
\caption{{\it Color online}. Ground state density maps for a two-dimensional mesoscopic system of $N$=19 dipolar bosons confined in a harmonic trap of  two different characteristic lengths $\xi$. All lengths are all in units of $a$ (see text). 
Left part ahow the non-classical crystal, while the right one the classical one. }\label{f4}
\end{figure}
\\ \indent
Let us now focus our attention on what could be characterized as the  mesoscopic analog of a {\it first-order} transition between a {\it non-classical crystalline ground-state} and the classical one. As shown in panels (c), (d) of Figure \ref{f2}, these crystals differ by the number of particles in the two rings, i.e., with respect to rotational symmetry, which is either ${\cal C}_4$ (panel (c)) or ${\cal C}_3$ (panel (d)). 
The same effect is observed in a cluster comprising $N$=19 particles (see Figure \ref {f4}). In this case, the non-classical crystal features eight particles in the inner ring, whereas the classical one has seven. For all other clusters with $N\le 30$, the only crystalline ground state is the classical one; it is worth noting that this fact has no consequence on the presence of a supersolid phase, which appears in any case at low $T$. However, for $N=12$ and $N=19$ the supersolid phase only occurs if the underlying crystalline phase is the non-classical one.
\\ \indent
In general, for a given number of particles $N$ there can be several low-energy classical equilibrium configurations, which differ from the ground state configuration in the number of particles belonging to each ring~\cite{Belousov00}.
For example, configuration ${\cal C}_4$ is one of these low-lying classically excited states for a system of $N$=12 particles; for $\xi \ll 1$,   ${\cal C}_3$ will  be the ground state, as the classical potential energy will overwhelm any contribution due to zero-point motion. On ther other hand, for intermediate values of $\xi$, the energy contribution associated to quantum displacement of particles around classical equilibrium positions stabilize ${\cal C}_4$ as the ground state.
\begin{figure}
\centering
\begin{tabular}{cc}
\includegraphics[width=0.46\columnwidth]{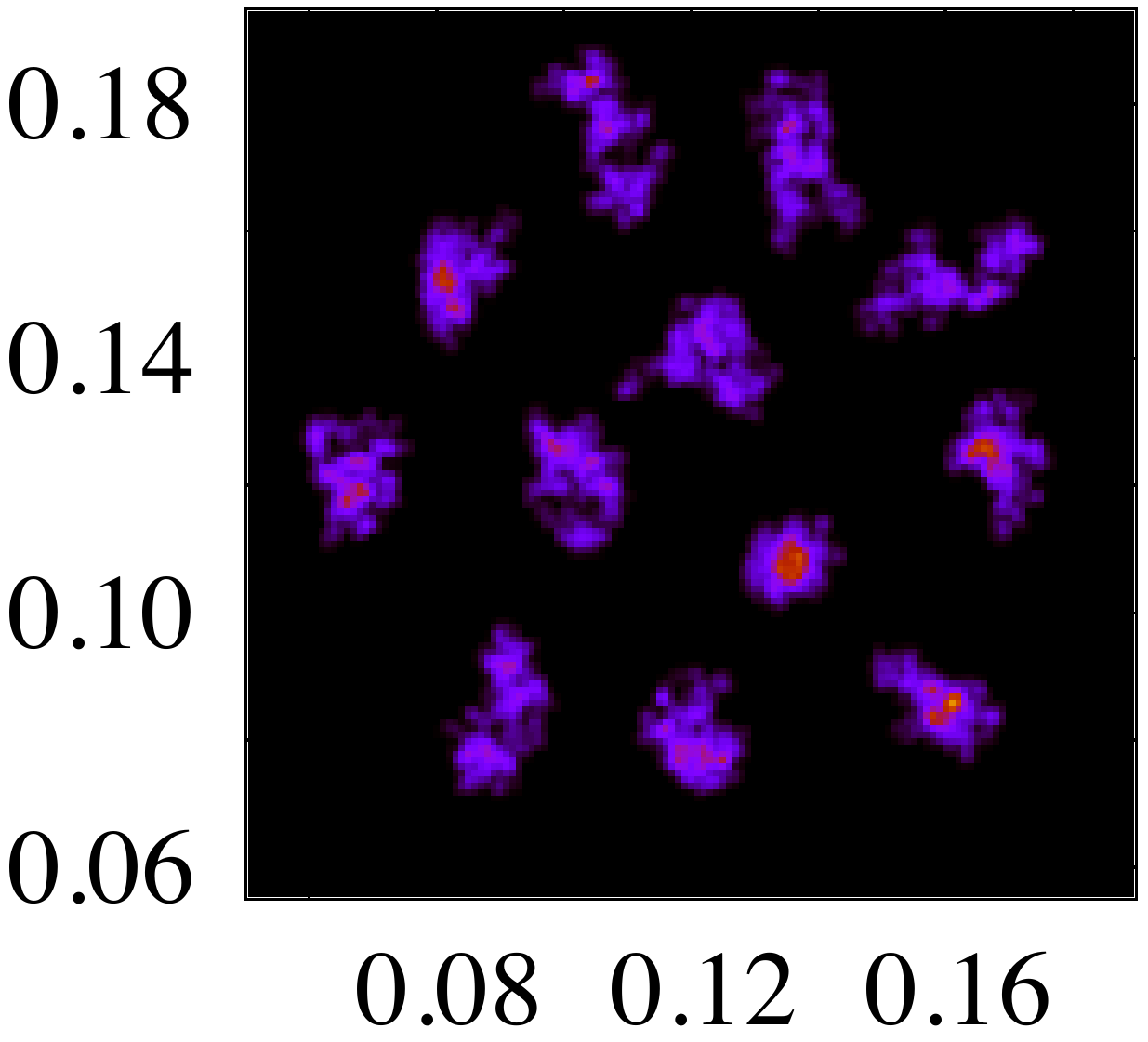}&
\includegraphics[width=0.46\columnwidth]{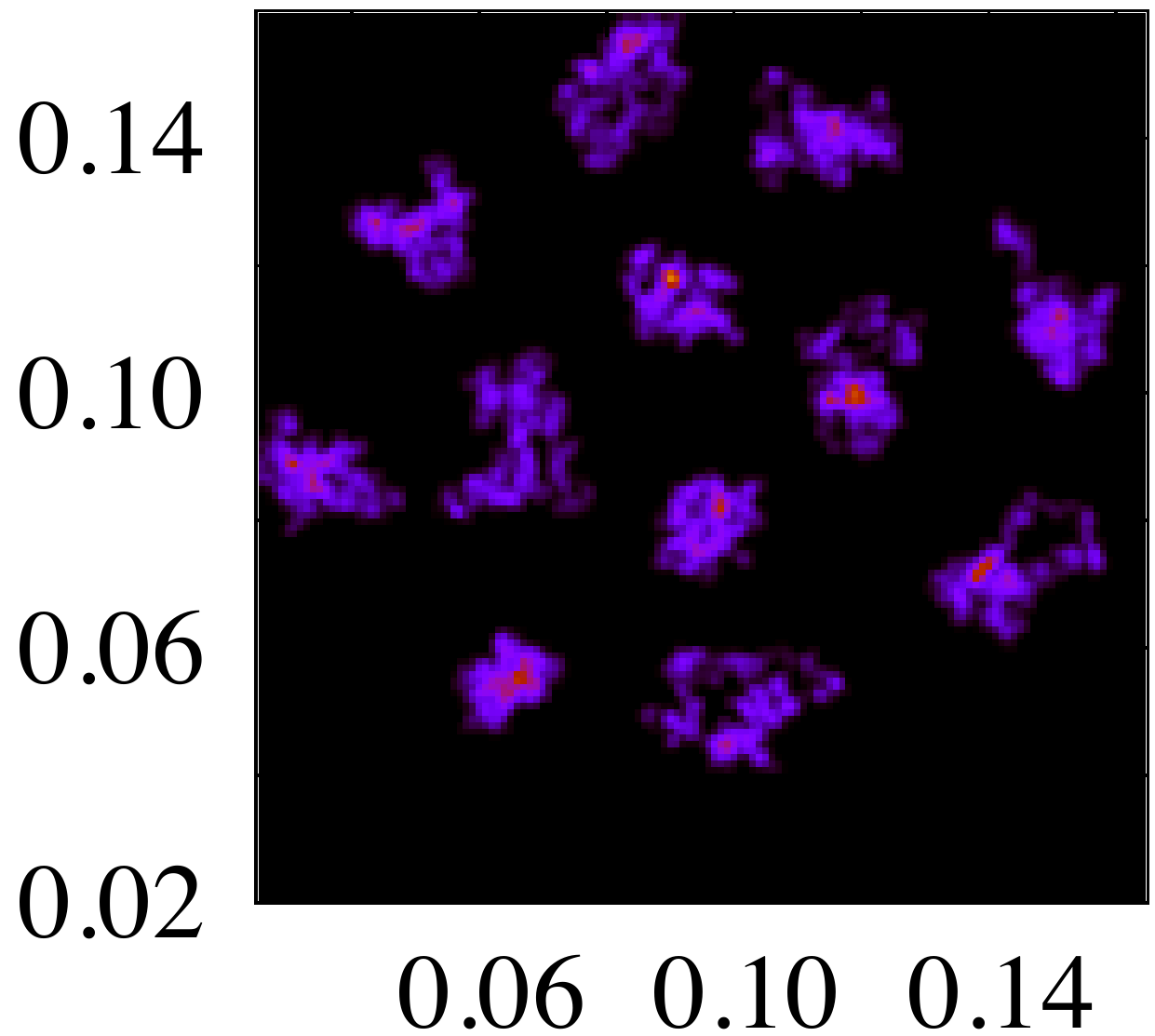}\\
\end{tabular}
\caption{{\it Color online}. Configurational snapshots for a two-dimensional mesoscopic system of $N$=12 dipolar bosons confined in a harmonic trap of  length $\xi=0.01$. All lengths are all in units of $a$ (see text). The temperature of the simulation is $T=2.4\times 10^3\ \epsilon_\circ$,  about thirty times higher than that which yields ground state physics. The system is found with roughly the same frequency in both  crystalline configurations shown here, namely ${\cal C}_3$ (left) and ${\cal C}_4$ (right). In the low temperature limit, ${\cal C}_4$ prevails, for this value of $\xi$. }\label{f3}
\end{figure}
\\ \indent
Most interestingly, we find that in the  range of values of $\xi$ in which the non-classical crystal (e.g., for $N$=12, ${\cal C}_4$) is the ground state, the classical one (${\cal C}_3$)  remains a low-temperature metastable conﬁguration. Specifically, in a finite range of temperature the system is observed in the Monte Carlo simulation to switch between ${\cal C}_4$ and ${\cal C}_3$. This is shown qualitatively in Figure \ref{f3}, displaying typical configuration snaphots yielded by a simulation carried out at a temperature $T=2.4\times 10^3\ \epsilon_\circ$; this is approximately thirty times higher than that yielding essentially ground state estimates, for which the configuration is that shown in the right panel of Figure \ref{f3}. The kinetic energy per particle at this high temperature is around $3T$. Melting  of the cluster into a featureless fluid takes place at a higher temperature, in fact at a value close to that of the bulk crystal \cite{Kalia81}. A similar coexistence is observed for the $N$=19 cluster as well.
\section{Discussion and Conclusions}
The physical behavior described above, namely the coexistence of two ``phases" of the cluster at finite temperature, with the classical one emerging as the temperature is raised, is analogous to what predicted for some small parahydrogen clusters (between 20 and 30 molecules) \cite{noi,noi2}. This intriguing type of phase coexistence is allowed by the finite size of the system, with the ensuing interplay between bulk and surface energy. An important difference is that, in the case of $p$-H$_2$ clusters, coexistence is between superfluid and crystalline phases, the superfluid being the ground state, underlain by particle exchanges. At higher temperature, as the thermal wavelength of the molecules becomes shorter, exchanges are suppressed and the system finds it energetically advantageous to take on solid-like. In the case of the dipolar cluster discussed here, coexistence is between two crystalline phases. Exchanges are infrequent in these phases, quantum-mechanical effects consisting almost exclusively of zero-point motion. An important consequence of this fact, is that a transition between non-classical and classical ground states can be expected to occur in mesoscopic dipolar systems of either Bose or Fermi particles. This is because any physical difference between Bose and Fermi systems can emerge only in the presence of quantum-mechanical exchanges of indistinguishable particles,  which are strongly suppressed in these mesoscopic crystals.
\\ \indent
In conclusion, a numerical studies of mesoscopic systems of dipolar particles confined to two dimensions yields evidence of non-classical crystalline ground states for clusters of specific numbers, underlain by quantum zero-point motion.
The mesoscopic phases described in this work  can be realized with polar molecules of current experimental interest. For example, for a moderate in-plane confinement $\omega / 2 \pi=$ 1kHz and fully polarized RbCs ($d=1.25$\ Debye), LiCs ($d=5.5$\ Debye), and SrO molecules ($d=8.9$ Debye), $\xi$ reaches values as small as  $\xi \simeq 0.031, 0.004$ and 0.002, respectively. Since the in-situ interparticle distances can be of the order of several hundreds of nm, it may be possible to directly address single particles {\it in situ}, and thus image the {\it spatial structure} of the crystalline phases above using, e.g., tightly focused beams. Alternatively, we propose the following method, which amounts to a version of a {\it magnifying lens}. At a given time $t_0$ the (DC or AC) fields inducing the dipole-dipole interactions are switched off, and the in-plane harmonic confinement is inverted in sign. Because of this inverted potential, each particle experiences a radial acceleration which depends on its {\em spatial position} at time $t_0$. After a certain time-of-flight, the particles can be, e.g, ionized and their positions recorded on a ion plate with unit efficiency, providing a magnified picture of the in-situ spatial configuration.
\\


\begin{thebibliography}{40}
\bibitem{qualifier} Obviously,  in a strict thermodynamic sense one cannot speak of ``phase'' of a finite system. Henceforth, therefore, this word should be interpreted loosely, in reference to qualitative, average properties of the system in its most probable configurations.
\bibitem{note} Obviously, any finite cluster can rotate without dissipation at a temperature lower than $\sim \hbar^2/I$, $I$ being the classical moment of inertia of the cluster.  Such a free rigid rotation constitutes a conceptually different scenario from the one referred to here as ``superfluid".
\end{thebibliography}
\end{document}